%% file: pasguide.tex
\documentclass{pas}

\usepackage{graphicx}
\usepackage{overpic}
\usepackage{xspace}
\usepackage{amssymb}
\usepackage{hyperref}
\usepackage{booktabs}
\usepackage{enumitem}
\usepackage{tabularx}

\def\ha{H$\alpha$}
\def\hb{H$\beta$}

\def\oiii{[O~{\sc iii}]}
\def\nii{[N~{\sc ii}]}

\def\hii{H~{\sc ii}}

\def\rre{\ensuremath{R/R_e}\xspace}
\def\sigmastar{\ensuremath{\Sigma_\star}}
\def\logsigmastar{\ensuremath{\log_{10}\left(\sigmastar\right)}\xspace}

\usepackage{microtype,siunitx,booktabs,multicol}
\makeatletter
\DeclareRobustCommand{\HI}{%
  \mbox{H\check@mathfonts\fontsize\sf@size\z@\selectfont I}%
}
\DeclareRobustCommand{\HII}{%
  \mbox{H\check@mathfonts\fontsize\sf@size\z@\selectfont II}%
}
\makeatother

\usepackage{multirow}

\begin{document}

\lefttitle{Publications of the Astronomical Society of Australia}
\righttitle{A. Mailvaganam \textit{et al.}}

\jnlPage{1}{13}
\jnlDoiYr{2026}
%\doival{10.1017/pasa.xxxx.xx}

\articletitt{Research Paper}

\title{Predicting Resolved Dust Attenuation from Local Galaxy Properties Using MaNGA}

\author{
\sn{Anilkumar} \gn{Mailvaganam}$^{1,2,3}$,
\sn{Tayyaba} \gn{Zafar}$^{1,2}$,
\sn{Pablo} \gn{Corcho-Caballero}$^{4,3}$,
\sn{Tamal} \gn{Mukherjee}$^{1,2,3}$,
\sn{Jahang} \gn{Prathap}$^{1,2,3}$,
\sn{Kyle B.} \gn{Westfall}$^{5}$,
\sn{Kevin} \gn{Bundy}$^{5}$
}

\affil{$^1$School of Mathematical and Physical Sciences, Macquarie University, NSW 2109, Australia}

\affil{$^2$Macquarie University Research Centre for Astronomy, Astrophysics \& Astrophotonics, Sydney, NSW 2109, Australia}

\affil{$^3$ARC Centre of Excellence for All Sky Astrophysics in 3 Dimensions (ASTRO-3D), Australia}

\affil{$^4$Kapteyn Astronomical Institute, University of Groningen, PO Box 800, 9700 AV Groningen, The Netherlands}

\affil{$^5$University of California Observatories, University of California, Santa Cruz, 1156 High Street, Santa Cruz, CA 95064, USA}

\corresp{A. Mailvaganam, anilkumar.mailvaganam@hdr.mq.edu.au}

%\citeauth{Author1 C and Author2 C, an open-source python tool for simulations of source recovery and completeness in galaxy surveys. {\it Publications of the Astronomical Society of Australia} {\bf 00}, 1--12. https://doi.org/10.1017/pasa.xxxx.xx}

\citeauth{Mailvaganam A et al., Predicting Resolved Dust Attenuation from Local Galaxy Properties Using MaNGA. {\it Publications of the Astronomical Society of Australia}.}

%\history{(Received xx xx xxxx; revised xx xx xxxx; accepted xx xx xxxx)}

\begin{abstract}
Accurate spatially resolved dust corrections are critical for interpreting the structure and evolution of star-forming galaxies (SFGs). We present an empirical model for predicting spatially resolved dust attenuation ($A_V$) in SFGs using integral field spectroscopy from the Mapping Nearby Galaxies at Apache Point Observatory (MaNGA) survey. Using a sample of 5,155 galaxies over $7.20<M_\ast<11.14$ and $0.0002 < z < 0.1444$, we derive $A_V$ maps from the Balmer decrement across more than 1,898,954 star-forming spaxels. Using local star formation rate surface density ($\Sigma_{\text{SFR}}$) as a predictor, the model achieves $R^2 = 0.69$ and RMSE $=0.22$ mag, with residuals that are approximately Gaussian and centred near zero. It predicts $A_V$ within a factor of $\sim$1.3 on kpc scales. We also demonstrate that the relation can be applied iteratively to recover dust–corrected $\Sigma_{\mathrm{SFR}}$ from uncorrected values, converging by the fourth iteration with minimal residual bias ($-0.01$ mag) and low RMSE ($0.42$ mag). The model accurately reproduces $A_V$ maps across diverse morphologies and orientations, including edge-on systems. It also recovers the observed radial $A_V$ profiles, capturing their dependence on stellar mass and relative star formation activity, with more massive and more strongly star-forming galaxies showing steeper gradients. 
\end{abstract}

%\begin{keywords}
%Key1, Key2, Key3, Key4
%\end{keywords}

\maketitle

\section{Introduction}

Dust particles make up about one percent of the material in interstellar space, yet they play a critical role in the formation of galaxies, stars, and planets \citep{draine03}. In galaxies, dust obscures ultraviolet (UV) and optical light and re-emits it in the infrared, making dust attenuation ($A_V$) a key observable for tracing star formation activity, the buildup of metals through chemical enrichment, and interstellar medium (ISM) geometry.

$A_V$ reflects the combined effect of dust extinction and geometry, and it varies across galaxies depending on local conditions \citep{witt92,calzetti94,draine11,tomicic17}. The intensity ratio of the Balmer lines (\ha/\hb\ or the so-called Balmer decrement (BD)) provides a measure of $A_V$ in the star-forming \hii\ regions, where a higher ratio indicates more $A_V$ with minimal temperature dependence \citep{baker38,osterbrock06,garn10,groves12,nelson16,greener20}. Observations show that $A_V$ correlates with both stellar mass surface density ($\Sigma_\ast$) and star formation rate surface density ($\Sigma_{\text{SFR}}$), indicating a strong link between dust, star formation, and underlying stellar structure \citep{Grootes_2013,Abdurro_28,Li_2021,Wild11}.

While previous studies relied on integrated spectra, recent advancements in integral field spectroscopy (IFS) enable spatially resolved attenuation measurements across galaxies \citep[e.g.,][]{Sanchez12,bundy15,Croom2021}. This allows for a pixel-by-pixel examination of how $A_V$ varies with local galaxy properties, revealing radial trends and variations tied to physical conditions such as $\Sigma_\ast$ and $\Sigma_{\mathrm{SFR}}$ \citep{Li_2024,Bluck_2019}.

Far-infrared (FIR) observations trace dust emission directly, as UV light from massive stars is absorbed by dust grains and re-emitted thermally. The shape of the FIR spectral energy distribution (SED) provides information about dust temperature, composition, and grain size \citep{Draine_2001, Li_2012, Jones_2023}. Surveys such as the SIRTF Nearby Galaxies Survey (SINGS; \citealt{Kennicutt_2003}) and the Key Insights on Nearby Galaxies: a Far-Infrared Survey with \textit{Herschel} (KINGFISH; \citealt{Kennicutt_2011}) provide spatially resolved FIR maps of nearby galaxies. However, FIR and submillimeter observations are limited by atmospheric opacity and the spatial resolution of space-based instruments, with no current or planned missions expected to achieve better than the 1-arcsecond barrier \citep{Rigopoulou_2021, Linz_2020, DENNY2013114}. Optically derived IFS extinction maps offer a complementary, higher-resolution approach for studying the internal distribution of dust across galaxies.

Radiative transfer (RT) modeling is often needed to convert physical dust properties into observables like $A_V$, as it captures how dust, starlight, and geometry interact. However, RT techniques are computationally expensive and sensitive \citep[e.g.,][]{Narayanan_2011, Narayanan_2021} to assumptions about dust geometry, grain properties, and the distribution of radiation sources \citep[see for a comprehensive review][]{Steinacker_2013}. Due to these challenges, many large-scale cosmological simulations adopt simplified attenuation prescriptions \citep{Camps_2016, Trayford_2017}, often scaling extinction with gas column density or metallicity while assuming fixed gas-to-dust ratios \citep{Trayford_2015, Nelson_2017}. These approximations lack the spatial resolution and flexibility to capture the observed variation of $A_V$ across galactic environments.

While the BD provides a physically motivated estimate of $A_V$, its practical application is often limited by observational constraints. H$\alpha$ is bright and relatively easy to detect; however, H$\beta$ is typically much fainter and more sensitive to noise. The H$\beta$ line often falls below detection limits, particularly in low surface brightness regions or at high redshifts, where it remains intrinsically weak and H$\alpha$ shifts out of the optical window \citep{Steidel_2014, Reddy_2015, Koyama_2015,garn10}. Even in deep spectroscopic surveys, the low signal-to-noise (SNR) of H$\beta$ continues to limit the applicability of BD–based attenuation estimates \citep{groves12, Li_2021}. Empirical approaches that predict $A_V$ using more readily available galaxy properties provide valuable alternatives \citep{Koyama_2015}. In this work, we use spatially resolved spectroscopy from the Mapping Nearby Galaxies at Apache Point Observatory (MaNGA) survey \citep{bundy15} to investigate how $A_V$ correlates with these local galaxy properties.

The paper is organised as follows. In Section~\ref{data}, we describe the MaNGA dataset used in this study. Section~\ref{methods} outlines our methodology, including the construction of $A_V$ maps, the derivation of local physical properties, and the sample selection. In Section~\ref{result}, we develop and evaluate our empirical models to predict $A_V$. Section~\ref{discussions} compares the best model to the observed spatial and radial $A_V$ distributions and discusses the physical implications. Finally, Section~\ref{conclusion} summarises our conclusions. Throughout the paper, we adopt a flat $\Lambda$CDM cosmology with $\Omega_\Lambda=0.7$, $\Omega_m=0.3$, and $H_0=70$ km s$^{-1}$ Mpc$^{-1}$.

\section{Data}
\label{data}

\subsection{The MaNGA survey}
\label{sec:manga}

The MaNGA survey is one of the three core projects of  Sloan Digital Sky Survey IV \citep[SDSS-IV;][]{Blanton_2017}, conducted using the 2.5 meter telescope at Apache Point Observatory \citep{Gunn_2006}. The survey targeted a total of $\sim10,000$ nearby galaxies across a redshift range of $0.01 < z < 0.15$, with a median redshift of $z\sim0.037$ \citep{Law_2016}. The galaxy sample was selected from the NASA-Sloan Atlas\footnote{\url{http://nsatlas.org/}}, a catalog combining Galaxy Evolution Explorer \citep[GALEX;][]{Martin_2005}, SDSS, and Two Micron All-Sky Survey \citep[2MASS;][]{Skrutskie_2006} photometry, and was designed to span a broad range of stellar masses ($M_\ast$), from $5 \times 10^{8}\,M_\odot$ to $3 \times 10^{11}\,M_\odot,h^{-2}$, with an approximately flat distribution in $M_\ast$ for galaxies in the local Universe \citep{Yan_2016, Wake_2017}. 

The MaNGA IFUs consist of hexagonal fibre bundles containing 19 to 127 fibres, providing fields of view from $12''$ to $32''$ in diameter, with each fibre covering $2''$ on the sky \citep{Law_2016}. The reconstructed datacubes have a spatial sampling of $0.5''$ per spaxel and a median point spread function (PSF) full-width at half-maximum (FWHM) of $\sim2.5''$, with a spectral resolution of $R\sim2,000$ covering a wavelength range of 3,600–10,300\,\AA . Data calibration and reduction are performed using the MaNGA Data Reduction Pipeline \citep[DRP;][]{Law_2016, Yan_2016_SCT}. For this work, we use $\Sigma_\ast$ and emission line maps derived with the \texttt{Pipe3D} pipeline (version 3.1.1; \citealt{Sanchez_2022}), which was applied to MaNGA DR17 \citep{Abdurro_uf_2022}. \texttt{Pipe3D} fits the stellar continuum with simple stellar population models, measures nebular emission lines across each spaxel, and applies a Galactic extinction correction using the \citet{Schlegel_1998} dust maps and the \citet{cardelli89} extinction law with $R_V = 3.1$.

\section{Methods}
\label{methods}

\subsection{Attenuation maps}\label{attenuation_maps}

In this study, we construct MaNGA $A_V$ maps from the sensitivity of the ${\rm BD}\equiv F(H\alpha)/F(H\beta)$, based on the emission line flux maps described in \S\ref{data}, to account for dust extinction effects. Since the emission line maps have already been corrected for foreground Galactic extinction by the \texttt{Pipe3D} pipeline (see \S\ref{data}), no additional Galactic correction was applied in this work. When interpreting \ha\ emission used for estimating $A_V$ maps, it is important to note that it can originate from various excitation mechanisms in galaxies, i.e., $i)$ from \hii\ regions dominated by star formation processes, $ii)$ through gas photoionisation by an active galactic nucleus (AGN; \citealt{trippe15}), or $iii)$ by collisional ionisation of interstellar shocks \citep{dopita76,kehrig12}. These mechanisms lead to different ionised gas conditions, resulting in distinct emission line ratios. In this work, we restrict the analysis to regions dominated by star formation, excluding regions where line excitation mechanisms may have different physical origins (e.g. AGN and shocks).

Therefore, to select relevant star-forming spaxels, we use the \nii/\ha~vs~\oiii/\hb\ Baldwin-Phillips-Terlevich \citep[BPT;][]{bpt81} diagnostic diagram. This diagnostic differentiates \hii\ regions from AGN- and shock-dominated regions distinguished by the high-excitation \oiii$\lambda$5007/\hb\ and low-excitation \nii$\lambda$6584/\ha\ line ratios. \citet[][hereafter Ke01]{Kewley_2001} established a theoretical demarcation for AGN and \hii\ regions, while \citet[][hereafter Ka03]{Kauffmann_2003b} provided an empirical classification scheme based on the study of $\sim$ 120,000 nearby SDSS galaxies. In the BPT diagram, pixels beneath the Ka03 line are considered purely excited by star formation, while those above the Ke01 line are interpreted as dominated by hard components (e.g., AGN). Based on this classification, we generate star formation masks to isolate spaxels solely dominated by star formation. We note that residual low-ionisation emission-line regions (LIERs) may persist in some low $\Sigma_{\text{SFR}}$ spaxels even after BPT selection \citep{Belfiore_2016}, which could mildly influence attenuation estimates in these regions.

We apply the star formation masks to the MaNGA H$\alpha$ and H$\beta$ emission line maps, which allows us to isolate the purely star-forming regions. Across the full MaNGA sample of 10,243 galaxies where both H$\alpha$ and H$\beta$ emission lines are detected, we select spaxels with SNR $>5$ in both lines. After applying this cut, 6,083 galaxies remain in the sample, each containing spaxels that satisfy the SNR threshold and are used to estimate the BD.

We denote the observed and intrinsic Balmer decrements as $\rm BD_{obs}$ and $\rm BD_{intrinsic}$, respectively. Canonical \hii\ regions lie in the range $2.725 \leq BD \leq 3.041$, corresponding to ISM conditions with temperatures ranging from 5,000 to 20,000\,K and electron densities between $10^{2}$ and $10^{6}$\,cm$^{-3}$ \citep{osterbrock06}.

Assuming a dust screen model between an emitting source and an observer, the attenuated flux of the source is given by:
 
\begin{equation}
    \label{eq:flux_extinction}
    F(\lambda) = F_o(\lambda) 10^{-0.4 A_\lambda} = F_o(\lambda) 10^{-0.4 A_V \eta(\lambda)}
\end{equation}

where $F_o(\lambda)$ corresponds to the intrinsic flux of the source, $A_\lambda$ is the wavelength-dependent attenuation expressed in magnitudes (mag), which can be parametrised in terms of the extinction in the $V$-band and an extinction law $\eta(\lambda)$, such that $\eta(V)=1$ \citep[e.g.][]{cardelli89, calzetti01}. The tight constraint between H$\alpha$ and H$\beta$ fluxes, in combination with Equation~\ref{eq:flux_extinction}, allows us to infer the value of $A_V$ as:

\begin{equation}
\label{eq:a_v_definition}
    A_\lambda = \frac{2.5 \, \eta(\lambda)}{\eta({H\beta}) - \eta({H\alpha})} \log_{10}\left(\frac{\rm BD_{obs}}{\rm BD_{intrinsic}}\right)
\end{equation}

We adopt $\rm BD_{intrinsic}=2.86$, corresponding to Case B recombination \citep{osterbrock1989} conditions with a gas temperature of $T=10^4$K and an electron density of $n_e=10^2$ cm$^{-3}$, as well as the \citet{calzetti01} dust attenuation law. We use the $A_V$-normalised form of the attenuation curve, written as $A_\lambda = \eta(\lambda)\,A_V$. We compute the $A_V$ maps using the following expression:

\begin{equation}
    A_V = 6.09 \cdot \log_{10}\left(\frac{\rm BD_{obs}}{\rm BD_{intrinsic}}\right)
\end{equation}

$\text{BD}_{\text{obs}}$ below the intrinsic threshold, whether caused by measurement uncertainties or variations in ISM conditions, leads to negative $A_V$ values and is excluded from the analysis. Applying this criterion reduces the sample to 5,155 galaxies within $0.0002 < z < 0.1444$, spanning $7.20<M_\ast<11.14$ and comprising 1,898,954 star-forming spaxels.

\begin{figure}[ht]
    \centering
    \includegraphics[width=\linewidth]{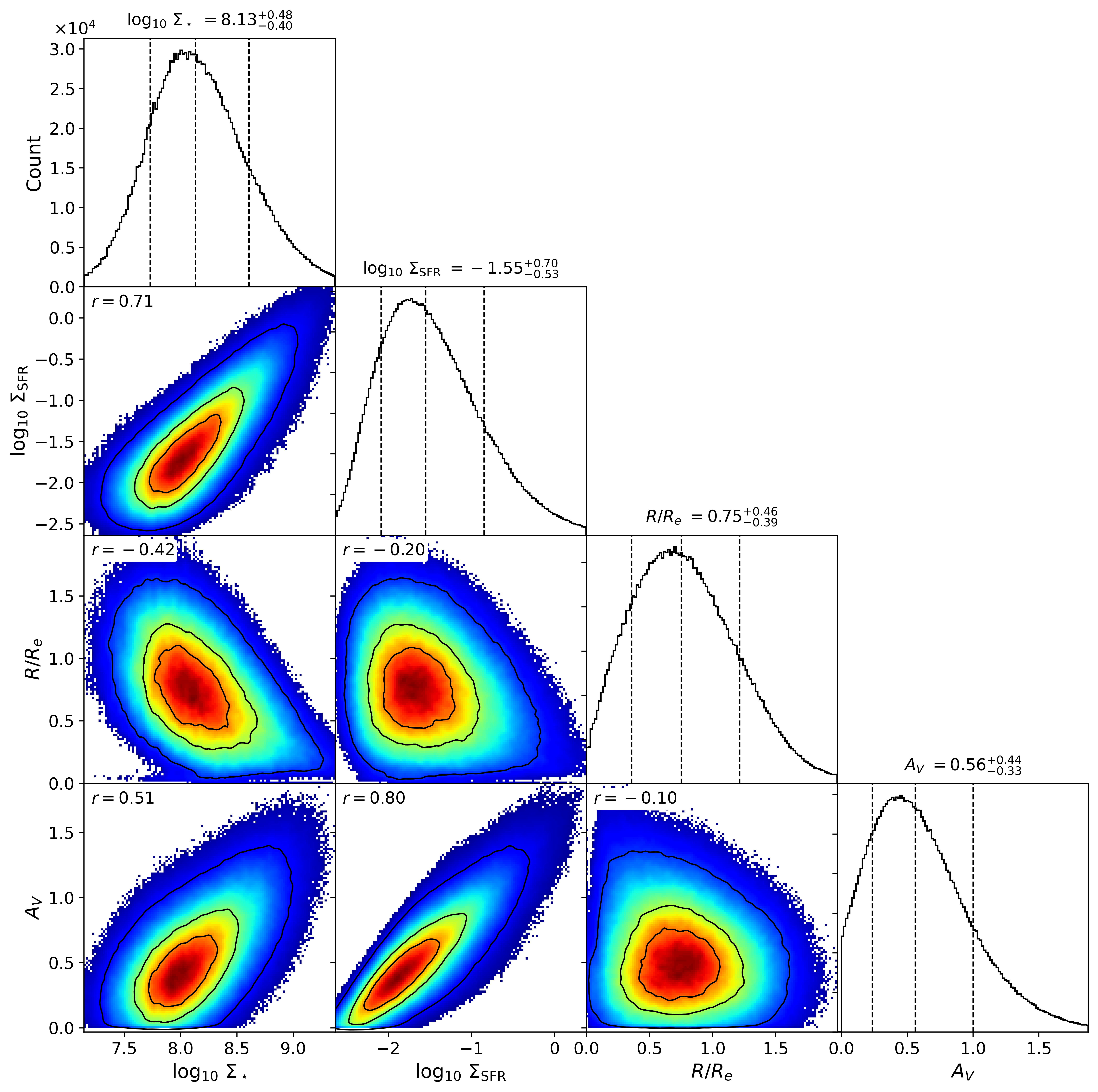}
    \caption{Corner plot showing the distributions and intrinsic relations between $\log_{10}\,\Sigma_\ast$, $\log_{10}\,\Sigma_{\mathrm{SFR}}$, $R/R_e$, and $A_V$ for star-forming spaxels in our sample. The diagonal panels display one-dimensional histograms with vertical lines marking the 16th, 50th, and 84th percentiles. The lower–triangle panels show the corresponding distributions with contours enclosing 25\%, 50\%, and 90\% of the data, with Spearman correlation coefficients $r$ indicated in each panel.}
    \label{fig:corner_plot}
\end{figure}

\begin{figure}[ht]
    \centering
    \includegraphics[width=8.5cm, height=6.5cm]{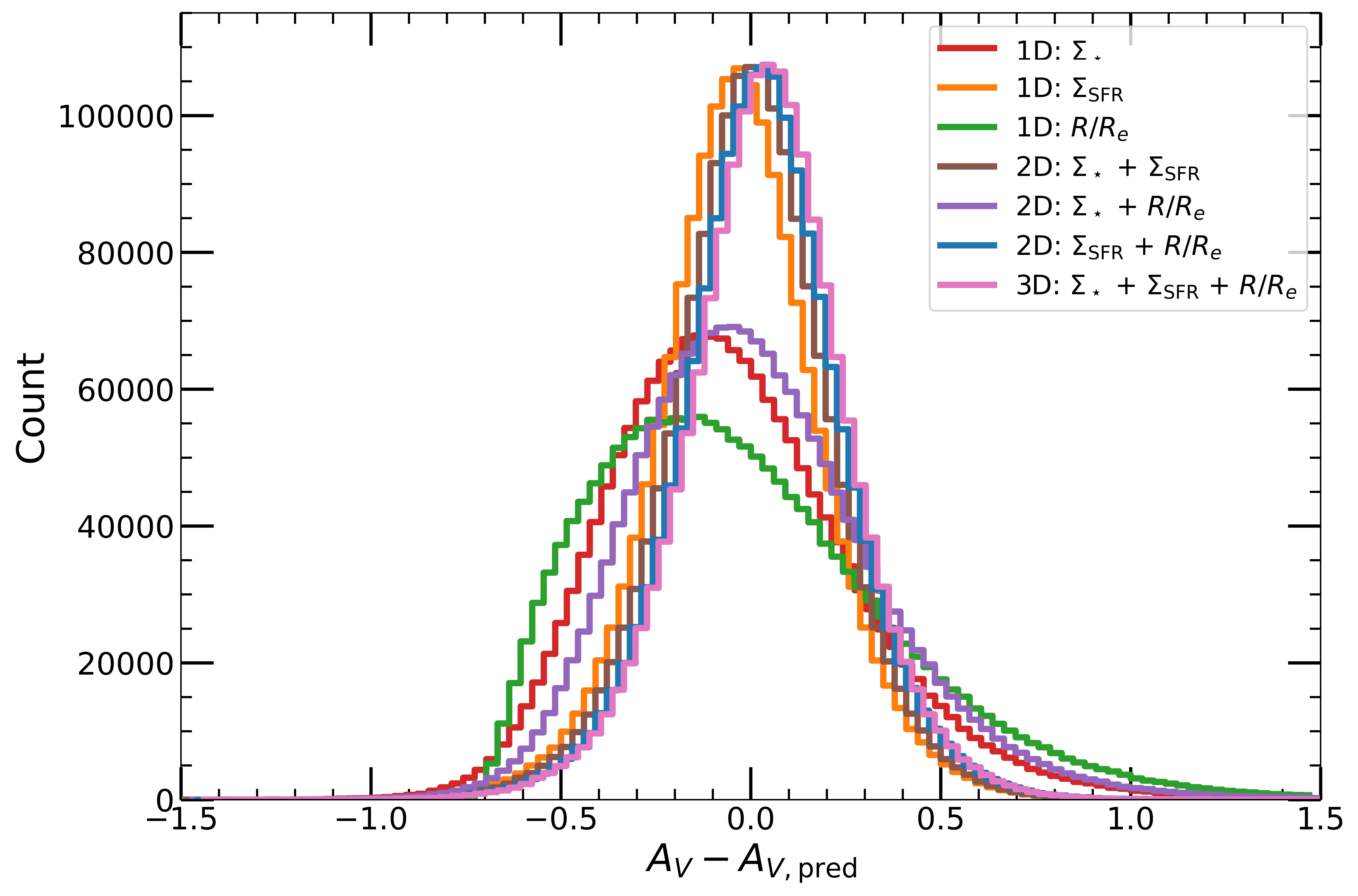}

    \caption{Residual distributions ($A_V - A_{V,\mathrm{pred}}$) for different OLS linear models. The x-axis shows residuals between observed and predicted $A_V$, and the y-axis shows the number of spaxels per bin. Each model, based on different combinations of predictors, is colour-coded.}
    \label{fig:2d_hist}
\end{figure}

\subsection{Stellar mass surface density maps}\label{stellar_mass_maps}

We use resolved $\Sigma_\ast$ maps derived from the full-spectral fitting performed by \texttt{Pipe3D} (see \citealt{Sanchez_2022} for a detailed description of the methodology). Each spaxel spectrum is modelled as a linear combination of Simple Stellar Population (SSP) templates, accounting for stellar kinematics and stellar continuum extinction using the extinction law of \citet{cardelli89}. The stellar mass-to-light ratio is estimated in the rest-frame V-band ($5,450$–$5,550$\,\AA\ ) as a weighted average of the SSP models. In low SNR regions, neighbouring spaxels are binned using a continuum segmentation algorithm \citep{Sanchez_2016_pipe3da}, and the derived stellar population parameters are redistributed to individual spaxels with the dezonification parameter \citep{Cid_Fernandes_2013}, preserving the spatial resolution of the maps. From these fits, \texttt{Pipe3D} computes $\Sigma_\ast$ by scaling the extinction-corrected V-band luminosity with the physical spaxel area and the galaxy's luminosity distance. The resulting maps are provided in units of $M_\odot\,\mathrm{arcsec}^{-2}$.

These maps are converted into physical units of $M_\odot\,\mathrm{kpc}^{-2}$ by applying the angular diameter distance ($D_{\rm A}$) in kpc/arcsec at each galaxy's redshift: 
\begin{equation}
\Sigma_\ast(M_\odot\,\mathrm{kpc}^{-2}) = \Sigma_\ast(M_\odot\,\mathrm{arcsec}^{-2}) \cdot \left(\frac{1}{D_{\rm A}}\right)^2
\end{equation}

\subsection{Star formation rate density maps}\label{SFR density maps}

We derive spatially resolved SFR maps from the \texttt{Pipe3D} \ha\ maps, corrected for nebular $A_V$ (using Equation~\ref{eq:flux_extinction} and \ref{eq:a_v_definition}). Only star-forming spaxels classified based on the BPT diagnostic (see \S\ref{attenuation_maps}) are retained in the SFR calculation. The SFR in each spaxel is calculated following \citet{Kennicutt98} calibration: 

\begin{equation}
 \text{SFR} = \frac{L({\rm H}\alpha)}{1.26\times10^{41}\,{\rm erg\,s^{-1}}} \,\rm  M_\odot\,yr^{-1}
\end{equation}

To convert from a Salpeter to a \citet{chabrier03} initial mass function (IMF), we divide the H$\alpha$ luminosities by 1.53 \citep{gunawardhana13, davies16}.

$\Sigma_{\mathrm{SFR}}$, in units of \(M_\odot\,\mathrm{yr}^{-1}\,\mathrm{kpc}^{-2}\), is computed as:
\begin{equation}
\Sigma_{\mathrm{\text{SFR}}}(M_\odot\,\mathrm{yr}^{-1}\,\mathrm{kpc}^{-2}) = \frac{\text{SFR}(M_\odot\,\mathrm{yr}^{-1})}{(0.5 \times D_{\rm A})^2}
\end{equation}
where \(0.5''\) is the MaNGA spaxel size and \(D_{\rm A}\) is the angular diameter distance in kpc/arcsec.

\begin{table*}[ht]
\centering
\caption{Summary of the OLS regression models predicting $A_V$ from different combinations of spatially-resolved galaxy properties: stellar mass surface density ($\Sigma_\ast$), star formation rate surface density ($\Sigma_{\mathrm{\text{SFR}}}$), and normalised galactocentric radius ($R/R_e$). The left block lists the fitted regression coefficients, while the right block reports model performance metrics: coefficient of determination ($R^2$), reduced chi-square (Red.\,$\chi^2$), root-mean-square error (RMSE), the 16th/50th/84th percentiles of the residuals, and the Kolmogorov–Smirnov (KS) statistic of residual normality.}
\label{tab:linear_models_ols_combined}
\resizebox{\textwidth}{!}{
\begin{tabular}{lrrrrccccccc}
\toprule
& \multicolumn{4}{c}{\textbf{Coefficients}} & \multicolumn{7}{c}{\textbf{Metrics}} \\
\cmidrule(lr){2-5} \cmidrule(lr){6-12}
\textbf{Model} & \textbf{Intercept} & $\boldsymbol{\Sigma_{\mathrm{SFR}}}$ & $\boldsymbol{\Sigma_\ast}$ & $\boldsymbol{R/R_e}$ & $\mathbf{R^2}$ & \textbf{Red.\,$\chi^2$} & \textbf{RMSE} & \textbf{P16} & \textbf{P50} & \textbf{P84} & \textbf{KS} \\
\midrule
$R/R_e$                                &  0.72 &   --   &   --   & -0.13 & 0.02 & 9.43 & 0.40 & -0.39 & -0.06 & 0.38 & 0.06 \\
$\Sigma_{\mathrm{SFR}}$                &  1.40 &  0.53 &   --   &   --   & 0.69 & 3.01 & 0.22 & -0.20 & 0.00 & 0.21 & 0.02 \\
$\Sigma_\ast$                         & -3.39 &   --   &  0.49 &   --   & 0.31 & 6.03 & 0.33 & -0.32 & -0.03 & 0.31 & 0.04 \\
\midrule
$R/R_e$ + $\Sigma_{\mathrm{SFR}}$      &  1.36 &  0.54 &   --   &  0.06 & 0.69 & 2.98 & 0.22 & -0.20 & 0.00 & 0.20 & 0.02 \\
$R/R_e$ + $\Sigma_\ast$               & -3.91 &   --   &  0.54 &  0.13 & 0.32 & 5.89 & 0.33 & -0.31 & -0.03 & 0.31 & 0.04 \\
$\Sigma_{\mathrm{SFR}}$ + $\Sigma_\ast$ &  2.41 &  0.59 & -0.11 &   --   & 0.70 & 3.00 & 0.22 & -0.20 &  0.00 & 0.20 & 0.02 \\
\midrule
$R/R_e$ + $\Sigma_{\mathrm{SFR}}$ + $\Sigma_\ast$ &  2.23 & 0.58 & -0.10 & 0.03 & 0.70 & 2.99 & 0.22 & -0.20 & 0.00 & 0.20 & 0.02 \\
\bottomrule
\end{tabular}
}
\end{table*}

\section{Results}\label{result}

\begin{figure*}
    \includegraphics[width=\linewidth]{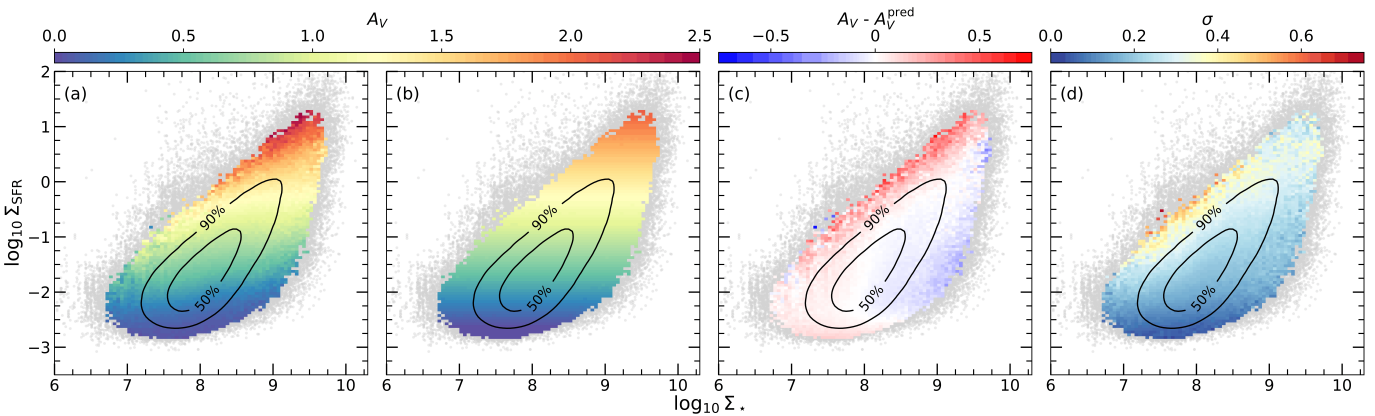}
    \caption{Distribution of $A_V$ over the $\log \Sigma_{\mathrm{SFR}}$-$\log \Sigma_\ast$ plane. Panels (a) and (b) show the observed and predicted (see Equation~\ref{eq:linear_av_from_sigma_sfr}) median $A_V$ values per bin (0.05 dex), respectively. Panels (c) and (d) display the median residuals ($A_V^\mathrm{obs} - A_V^\mathrm{pred}$) and associated standard deviation, $\sigma\left(A_V^\mathrm{obs} - A_V^\mathrm{pred}\right)$, respectively. Black contours enclose 90\% and 50\% of the sample.}
    \label{fig:av_obs_modelled}
\end{figure*}

\subsection{Developing an Empirical Model for Resolved Dust Attenuation}\label{av_model}

We first quantify the correlation of $A_V$ with $\log_{10} \Sigma_\ast$, $\log_{10} \Sigma_{\mathrm{SFR}}$, and the normalised galactocentric radius ($R/R_e$). Fig.~\ref{fig:corner_plot} illustrates the distributions of these parameters together with their interdependencies. These quantities can be derived directly from spatially resolved IFS data and reflect the physical interplay between stellar structure, star formation activity, and dust content that shape the spatial variation of $A_V$. Using Spearman rank correlation coefficients, we find that $\log_{10} \Sigma_{\mathrm{SFR}}$ shows the strongest positive correlation with $A_V$ ($r = 0.80$), followed by $\log_{10} \Sigma_\ast$ ($r = 0.51$). In contrast, $R/R_e$ exhibits only a weak negative correlation ($r = -0.11$). These results confirm that local $\Sigma_{\mathrm{SFR}}$ is the dominant predictor of $A_V$, with $\Sigma_\ast$ providing a secondary contribution and $R/R_e$ playing a negligible role, consistent with previous findings \citep{Bluck_2019}.

In this section, we develop and assess an empirical model to predict $A_V$ on a spaxel-by-spaxel basis using resolved galaxy properties from the MaNGA survey. Our aim is to provide a computationally efficient method for estimating $A_V$ from readily available local properties, offering an alternative in cases where BD measurements are unavailable or unreliable, or in simulations that provide only physical properties rather than direct observables. To construct the multi-dimensional model, we employ ordinary least squares (OLS) regression, exploring all combinations of three input parameters: $\log_{10} \Sigma_\ast$, $\log_{10} \Sigma_{\mathrm{SFR}}$, and $R/R_e$. The general form of the model is:

\begin{equation}
\label{eq:linear_model}
A_V = \beta_0 + \sum_i \beta_i x_i
\end{equation}Here $x_i$ represents the predictor variables $\log_{10} \Sigma_\ast$, $\log_{10} \Sigma_{\mathrm{SFR}}$, and $R/R_e$, and the $\beta$ terms are the regression coefficients estimated from the fit of the data, restricted to spaxels within $R/R_{e} < 2.5$. Model performance is evaluated using the coefficient of determination ($R^2$), root mean square error (RMSE), and the Kolmogorov–Smirnov (KS) statistic to assess goodness of fit and residual distribution.

It is important to consider the intrinsic correlations among the predictor variables (see Fig.~\ref{fig:corner_plot}). $\log_{10}\Sigma_{\rm SFR}$ and $\log_{10}\Sigma_\ast$ are strongly correlated (Spearman $r = 0.71$) and both show clear radial trends with $R/R_e$, introducing redundancy among the predictors. Therefore, adding $\Sigma_\ast$ or $R/R_e$ to $\Sigma_{\rm SFR}$ provides limited additional information, which explains why multi-parameter models offer minimal improvements over single-parameter fits.

For each combination of the aforementioned parameters, the resulting model regression coefficients and associated performance metrics are summarised in Table~\ref{tab:linear_models_ols_combined}.

\begin{enumerate}[label=\roman*)]
    \item Single-parameter models show a broad range of performance. As expected, $\Sigma_{\mathrm{SFR}}$ performs best ($R^2=0.69$, RMSE = 0.22), followed by $\Sigma_\ast$ ($R^2=0.31$, RMSE = 0.33), while $R/R_e$ alone provides minimal predictive power ($R^2=0.017$, $\rm RMSE=0.40$). These results highlight $\Sigma_{\mathrm{SFR}}$ as the most effective single predictor of $A_V$.

    \item Two-parameter models provide only slight improvements over the single-parameter fits. The combination of $\Sigma_{\mathrm{SFR}}$ and $\Sigma_\ast$ performs best ($R^2=0.70$, RMSE = 0.219), closely followed by $\Sigma_{\mathrm{SFR}} + R/R_e$ ($R^2=0.69$, RMSE = 0.22), whereas $\Sigma_\ast + R/R_e$ produces only a weak relation ($R^2=0.32$, RMSE = 0.33). These results indicate that while $\Sigma_\ast$ offers some secondary predictive power by tracing metal-rich, gas-dense regions conducive to dust growth \citep{Zhukovska_2008, Draine_2009, Ludwig_2022}, $\Sigma_{\mathrm{SFR}}$ remains the primary driver of attenuation.

    \item Inclusion of $R/R_{e}$ in a three-parameter model (\logsigmastar, $\log_{10} (\Sigma_{\mathrm{SFR}})$, and \rre) does not significantly improve the fit compared to the two-parameter $\Sigma_{\mathrm{SFR}} + \Sigma_\ast$ model. Both achieve nearly identical coefficients of determination ($R^2=0.70$) and RMSE values (0.22 mag), and their residual distributions exhibit no significant variation. This confirms that $R/R_{e}$ provides no additional predictive power beyond what is captured by $\Sigma_\ast$ and $\Sigma_{\mathrm{SFR}}$.
    \end{enumerate}

We explore second-order (quadratic) extensions of all linear models for each combination of $\log_{10}\Sigma_{\mathrm{SFR}}$, $\log_{10}\Sigma_\ast$, and $R/R_e$. These quadratic fits produce only minimal improvements ($\Delta\mathrm{RMSE} \lesssim 0.002~\mathrm{mag}$; $\Delta R^{2} \approx +0.004$), and their residual distributions are nearly unchanged from the linear case. Given the minimal improvement and added model complexity, we keep the linear relation as our adopted model.

The $\Sigma_{\mathrm{SFR}}$ fit achieves a level of accuracy comparable to the best multi-parameter models, with $R^2=0.69$ and RMSE $=0.22$~mag. The residuals are approximately Gaussian and centred around zero (KS statistic $=0.022$), with a 1$\sigma$ interval of $[-0.20,\,+0.21]$\,mag. This corresponds to predicting $A_V$ to within a factor of $\sim$1.3. The final adopted linear model, fitted via OLS regression, has the form:

\begin{equation}
\label{eq:linear_av_from_sigma_sfr}
A_V = 1.40 \;+\; 0.53\,\log_{10}\Sigma_{\mathrm{\text{SFR}}}
\end{equation}

%In Fig.~\ref{fig:av_obs_modelled}, we compare the observed and predicted values of $A_V$ across the $\Sigma_{\mathrm{SFR}}$–$\Sigma_\ast$ plane. The parameter space is divided into bins of width 0.05 dex along both axes, with bins containing more than 25 spaxels colour-coded by the median $A_V$. Black contours enclose 50\% and 90\% of the total spaxel distribution, roughly tracing the resolved star-forming main sequence (SFMS; e.g. \citealt{Cano16, Abdurro_28, Sanchez_2020}).

\subsection{Observed vs. Modelled Dust Attenuation}\label{observed_vs_modelled}

In Fig.~\ref{fig:av_obs_modelled}, we compare the observed and predicted values of $A_V$ across the $\Sigma_{\text{SFR}}$–$\Sigma_\ast$ plane. The parameter space is divided into bins of width 0.05 dex along both axes, with bins containing more than 15 spaxels colour-coded by the median $A_V$. Black contours enclose 50\% and 90\% of the total spaxel distribution, roughly corresponding to the resolved star-forming main sequence (SFMS; e.g. \citealt{Cano16, Abdurro_28, Sanchez_2020}).

The observed $A_V$ distribution (panel (a) of Fig.~\ref{fig:av_obs_modelled}) shows a clear trend of increasing attenuation with higher $\Sigma_{\text{SFR}}$ and a weaker dependence on $\Sigma_\ast$. The strong dependence on $\Sigma_{\text{SFR}}$ is consistent with nebular emission predominantly tracing dusty \hii\ regions embedded within molecular clouds \citep{Battisti_2016, garn10, Qin_2023}. The colour gradient highlights the combined influence of these two quantities in shaping the local $A_V$ distribution. In particular, attenuation rises more steeply along the $\Sigma_{\text{SFR}}$ axis, reflecting our earlier finding that $\Sigma_{\text{SFR}}$ is the strongest single predictor of $A_V$ at the spaxel level. A weaker but noticeable dependence on $\Sigma_\ast$ is also present, indicating that $\Sigma_\ast$ contributes secondarily to the dust regulation.

Similar dependencies of stellar mass density and star formation activity have been shown to regulate star formation patterns within disks \citep{Gonzalez_Delgado_2016}, suggesting that similar mechanisms may also regulate the attenuation distribution. In our sample, however, the steeper gradient of $A_V$ with $\Sigma_{\mathrm{SFR}}$ compared to $\Sigma_\ast$ indicates that star formation activity is the primary driver. This motivates our choice of a 1D model based solely on $\Sigma_{\text{SFR}}$, which we next test against the observed attenuation distribution.

The predicted $A_V$ distribution (Fig.~\ref{fig:av_obs_modelled}; panel (b)) based on our empirical model successfully reproduces the main trends in the observed data, capturing both the overall gradient of $A_V$ and the regions of peak attenuation. As can be seen in panel (c), minor systematic residuals remain, with redder bins appearing along the upper envelope of the distribution ($7.3 \leq \log\Sigma_\ast \leq 9.3$ at high $\Sigma_{\mathrm{SFR}}$), indicating slight underestimation of $A_V$ in these extreme regimes. A similar trend of overestimation is visible at high $\Sigma_\ast$ and low $\Sigma_{\text{SFR}}$, but both effects lie outside the 90\% contour and involve only a small fraction of the spaxels.

\begin{figure*}[ht]
    \centering
    \includegraphics[width=\linewidth]{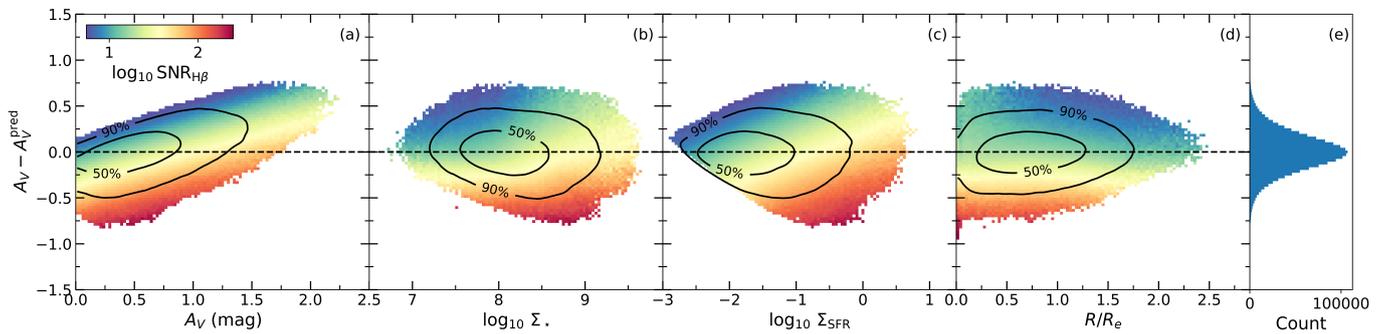}
    \caption{Residuals of the predicted $A_V$ from the empirical model as a function of (from left to right) observed $A_V$, $\log_{10} \Sigma_\ast$, $\log_{10} (\Sigma_{\mathrm{SFR}})$, and normalised galactocentric radius ($R/R_e$). In each panel, colours indicate the median $\log_{10}$ H$\beta$ SNR in each 2D bin, with 50\% and 90\% spaxel density contours overlaid in black. Horizontal dashed lines mark zero residual. The rightmost panel shows the overall residual distribution as a histogram.} 
    \label{fig:residual_plot}
\end{figure*}

In the densest regions of the $\Sigma_\ast$–$\Sigma_{\text{SFR}}$ plane (inside the contour enclosing 50\% of the spaxels), residuals are approximately symmetric and centred around zero, with a median of $-0.01$ mag and scatter $\sigma \approx 0.19$ mag, indicating that the model is well calibrated and qualitatively captures the dominant attenuation trends (see Fig.~\ref{fig:2d_hist}). Within the 90\% contour, residuals remain centred (median $=0.00$ mag) with scatter $\sigma \approx 0.20$ mag. Spaxels outside this region indicate a slight positive offset (median $=+0.03$ mag) and broader scatter ($\sigma \approx 0.24$ mag).

We examined whether incorporating the spaxel offset from the resolved SFMS ($\Delta \mathrm{sSFR}$) could improve the model, given the residual pattern visible in Fig.~\ref{fig:av_obs_modelled} (c). We computed $\Delta \mathrm{sSFR}$ using the median $\Sigma_\ast$–$\Sigma_{\text{SFR}}$ relation in our sample and refit the model with $\Delta \mathrm{sSFR}$ as an additional predictor. The resulting fit shows only a slight improvement ($\Delta \mathrm{RMSE} \approx 0.002$\,mag, $\Delta R^{2} \approx +0.007$), and the residual structure across the $\Sigma_\ast$–$\Sigma_{\text{SFR}}$ plane remains effectively unchanged. We therefore conclude that deviations from the resolved SFMS do not provide a significant improvement in predicting $A_V$ beyond what is already captured by $\Sigma_{\mathrm{SFR}}$.

\citet{Cano_Diaz_2019} showed that resolved studies of the SFMS can be biased by detection limits, sample selection, and radial aperture effects, particularly at low $\Sigma_\ast < 3 \times 10^7\, M_\odot\, \mathrm{kpc}^{-2}$, where the $\Sigma_\ast$–$\Sigma_{\mathrm{SFR}}$ relation may appear artificially flattened. Other MaNGA-based studies, however, have recovered resolved scaling relations without applying such a cut \citep[e.g.,][]{Omori_2022, Hsieh_2017}. In our analysis, we do not impose a low-$\Sigma_\ast$ threshold and find that our model predicts $A_V$ reliably in these regions, with no indication of the biases reported for the SFMS. This demonstrates that our empirical approach is robust across the full range of $\Sigma_\ast$ covered in our sample.

\subsection{Residual Trends and Systematics}\label{residual}

To assess potential biases in our empirical model, given by Equation~\ref{eq:linear_av_from_sigma_sfr}, we show in Fig.~\ref{fig:residual_plot} how the residuals between the observed and predicted $A_V$ vary with key physical parameters. We plot the distribution of residuals as a function of $A_V$ (a), $\log_{10}\Sigma_\ast$ (b), $\log_{10}\Sigma_{\mathrm{SFR}}$ (c), and $R/R_e$ (d), colour-coding each panel bin by the median H$\beta$ SNR with spaxels containing more than 25 spaxels.

Residuals as a function of observed $A_V$ remain broadly centred around zero (see also panel (e)), with only a mild positive correlation where the model slightly overpredicts at low $A_V$ and underpredicts at the highest attenuations. Within the 90\% contour, residuals roughly lie between $\pm0.5$ mag, while within the 50\% contour, they remain below $\pm0.25$ mag, indicating that the bulk of spaxels are well predicted. Deviations at very low and very high $A_V$ arise from comparatively few spaxels, reflecting limited statistics at the extremes of the distribution that are not well constrained by the OLS fit.

While the residual contours illustrate the overall distribution, we additionally performed a quantitative calibration check to assess the reliability of the model. Specifically, we computed the fraction of spaxel residuals falling within a fixed 90\% prediction interval ([-0.36,\,+0.37] mag). The coverage remains high ($\geq$ 90\%) for low-to-moderate attenuations ($A_V \lesssim 0.9$ mag) but declines steadily at higher values, falling to roughly 65\% around $A_V$ $\sim$1.5 mag and dropping below 20\% beyond 2.2 mag. This indicates that the model is well calibrated for low-to-moderate $A_V$, but its predictive confidence degrades systematically in dustier regions, reflecting the limited number of high $A_V$ spaxels available for calibration.

The residuals remain approximately flat when examined as a function of $\log_{10}\Sigma_\ast$, $\log_{10}\Sigma_{\mathrm{SFR}}$, and $R/R_e$, with no clear secondary trends. In all three cases, the median residual stays close to zero and the scatter remains modest, typically within $\pm 0.5$ mag for the central 90\% of spaxels. Apart from the mild correlation with observed $A_V$ discussed earlier, the absence of systematic residual trends across these other axes supports the reliability of the model and suggests that any remaining deviations are dominated by statistical scatter rather than physically correlated processes.

For comparison, residual trends for the 2D and 3D models are presented in the Appendix.

The colour gradient reveals a systematic dependence of residuals on H$\beta$ SNR, with positive residuals more common in spaxels with low SNR and negative residuals more frequent at higher SNR. This pattern likely arises because spaxels at the extremes of the H$\beta$ SNR distribution contribute little statistical weight to the regression, leaving the model less well constrained in those regimes. In the bulk of the distribution, the H$\beta$ SNR has a median of 18.3 (1$\sigma$ spread 12.0–28.4) inside the 50\% contour, rising slightly to 20.2 (1$\sigma$ spread 11.1–38.8) inside the 90\% contour. By contrast, spaxels outside the 90\% contour exhibit much broader and more heterogeneous SNR values, with a median of 41.3 and a 1$\sigma$ interval of 8.7–113.6. This indicates that the apparent residual trends at the extremes arise from sparsely sampled regions rather than the bulk of the data. A similar distribution is obtained when using H$\alpha$ SNR, but we adopt H$\beta$ SNR as the reference since it is the limiting line for the BD and therefore more directly constrains the attenuation estimates.

To assess whether this SNR-dependent behaviour arises from measurement quality effects rather than a physical correlation, we repeated the regression using weighted least squares (WLS), applying (i) inverse-variance weights ($1/\sigma_{A_V}^{2}$) and (ii) direct H$\beta$ SNR weighting. In both cases, the weighted fits produce only small shifts in the best-fit coefficients and slightly weaker performance relative to the unweighted OLS model. This indicates that the residual trend does not originate from SNR-driven measurement biases, but reflects intrinsic physical differences within the spaxel population.

\begin{figure*}[ht]
    \centering
    \includegraphics[width=\linewidth]{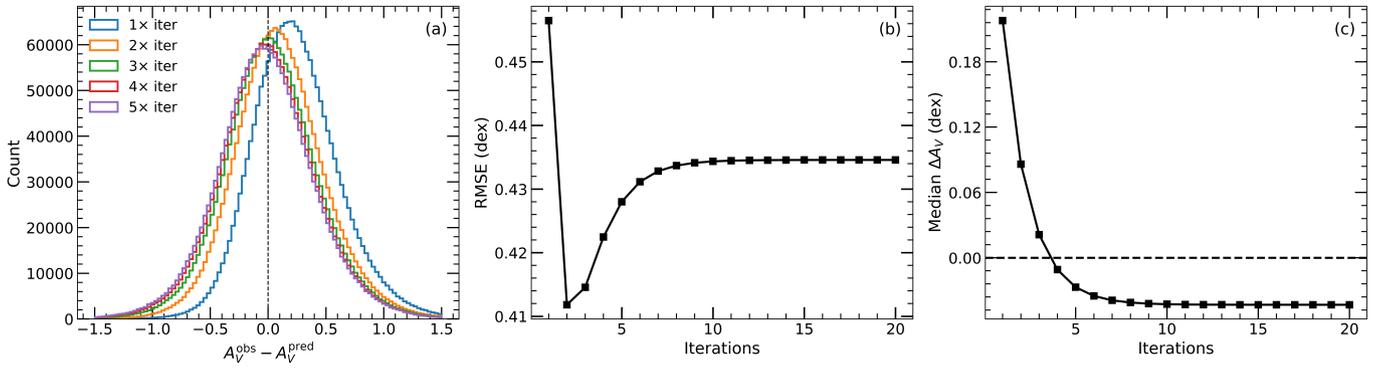}

    \caption{Residuals of the predicted $A_V$ from the iterative empirical correction. From left to right: (1) distributions of residuals for the first five iterations, (2) RMSE of the residuals as a function of iteration number, and (3) median residual offset as a function of iteration. Dashed vertical and horizontal lines indicate zero residual.}
    \label{fig:av_iterations}
\end{figure*}

\subsection{Iterative Recovery of Dust-Corrected SFR Surface Density}\label{dust_correction}

%%% Paragraph describing the algorithm

A key strength of our approach is that the empirical relation between $A_V$ and $\Sigma_{\mathrm{SFR}}$ can also be applied when starting from values of $\Sigma_{\mathrm{SFR}}$ derived from attenuated H$\alpha$ fluxes (Equation~\ref{eq:flux_extinction}) by adopting an iterative procedure. In the first iteration, a rough estimate of $A_V$, based on Equation~\ref{eq:linear_av_from_sigma_sfr}, is used to correct $\Sigma_{\mathrm{SFR}}$, which is then used to re-estimate $A_V$.
This process continues until the difference between the $A_V$ values obtained in two consecutive iterations remains below a given threshold ($|A_V^{(n)} - A_V^{(n-1)}| < \epsilon$).
In our case, we adopt $\epsilon=0.04$ as our stopping criterion, chosen to reflect the iteration plateau where further iterations produce negligible improvements.

%%% Paragraph describing the accuracy of the method

We assessed the iterative recovery of dust–corrected $\Sigma_{\mathrm{SFR}}$ across 20 iterations (Fig.~\ref{fig:av_iterations}). A single iteration produced a large median offset of $+0.22$ mag and the weakest agreement ($R^2 = 0.56$, RMSE = 0.46). By the second iteration, the residual offset was reduced to $+0.09$ mag, with $R^2 = 0.64$ and RMSE = 0.41, representing a strong improvement. Applying our stopping criterion, the procedure converges by the fourth iteration, at which point the residual offset is minimised ($-0.01$ mag), the RMSE remains low ($0.42$), and the fit maintains a strong agreement ($R^2 = 0.63$). Further iterations progressively reduce the median residual toward $\sim-0.04$ mag by iteration 10, but without further improvement in scatter ($\sigma = 0.40$ mag) or RMSE = $0.44$ mag). Therefore, iterations two through four all provide robust corrections, with the fourth iteration providing the smallest overall bias without degrading the overall fit performance.

\begin{figure}[ht]
    \centering
    \includegraphics[width=\linewidth]{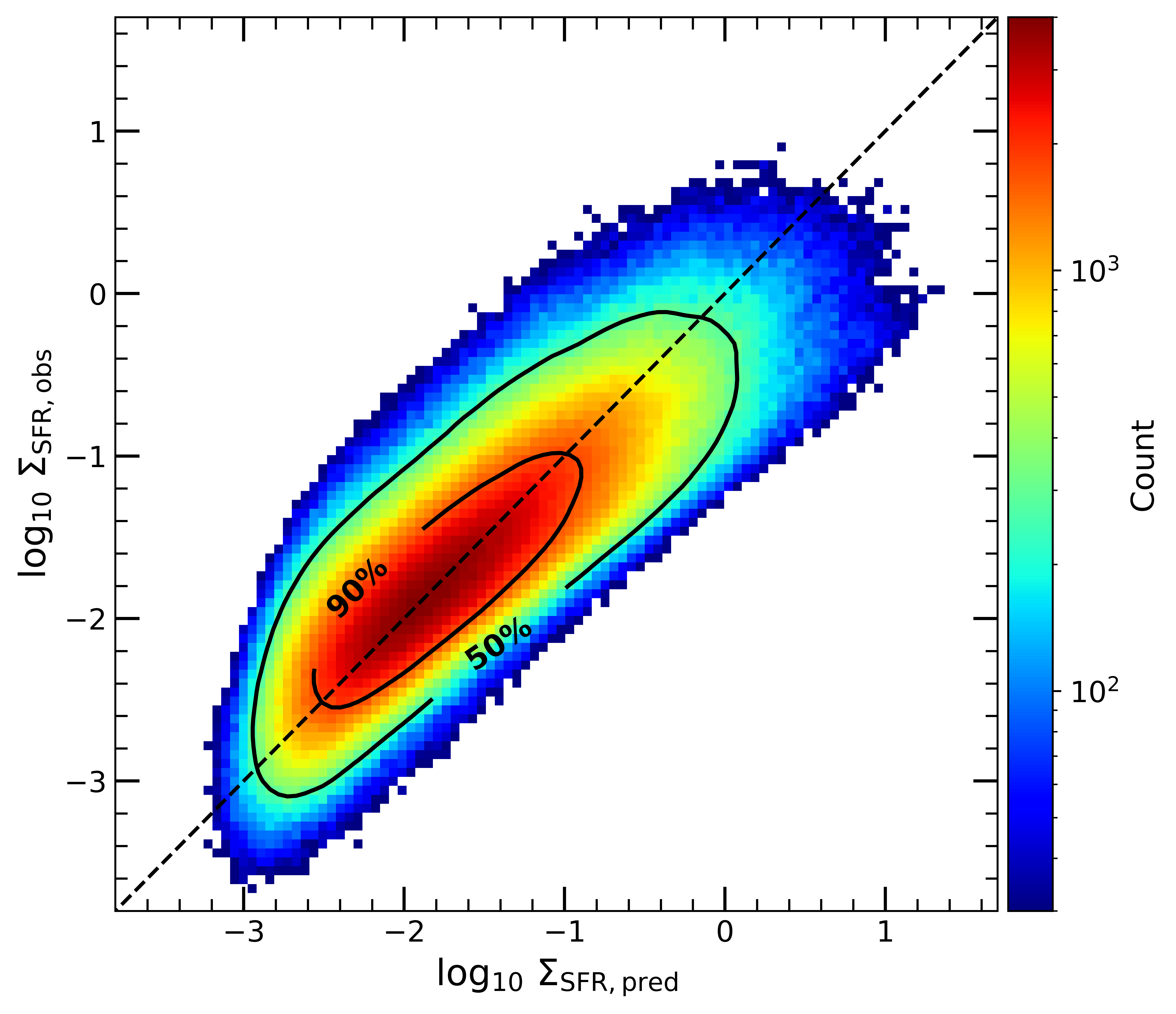}

    \caption{Comparison between predicted and Balmer–decrement corrected $\Sigma_{\mathrm{SFR}}$ after four iterations of the empirical attenuation relation. The colour map shows the spaxel density in the 2D histogram, with only bins containing at least 15 spaxels displayed. The dashed line indicates the one-to-one relation, and black contours enclose 50\% and 90\% of the spaxels.}
    \label{fig:sfr_pred}
\end{figure}

Fig.~\ref{fig:sfr_pred} illustrates the comparison between the recovered $\Sigma_{\mathrm{SFR}}$ after applying our iterative attenuation relation and the values corrected using the BD. The distribution closely follows the one-to-one line, with residuals centred near zero and the majority of spaxels enclosed within the 90\% contour. The residuals have a median of $-0.01$~dex, a typical 1$\sigma$ scatter of $0.39$~dex, and an RMSE of $0.42$~dex, indicating good overall recovery performance. This demonstrates that $A_V$ corrections can be obtained without direct reliance on H$\beta$ fluxes, enabling the recovery of corrected $\Sigma_{\mathrm{SFR}}$ in datasets where BD measurements are unavailable or uncertain.

%%%%%%%%%%%%%%%%%%%%%%%%%%%%%%

\section{Discussion}\label{discussions}

\begin{figure*}[t!]
    \centering
    \includegraphics[width=0.8\linewidth, height=0.9\textheight]
    {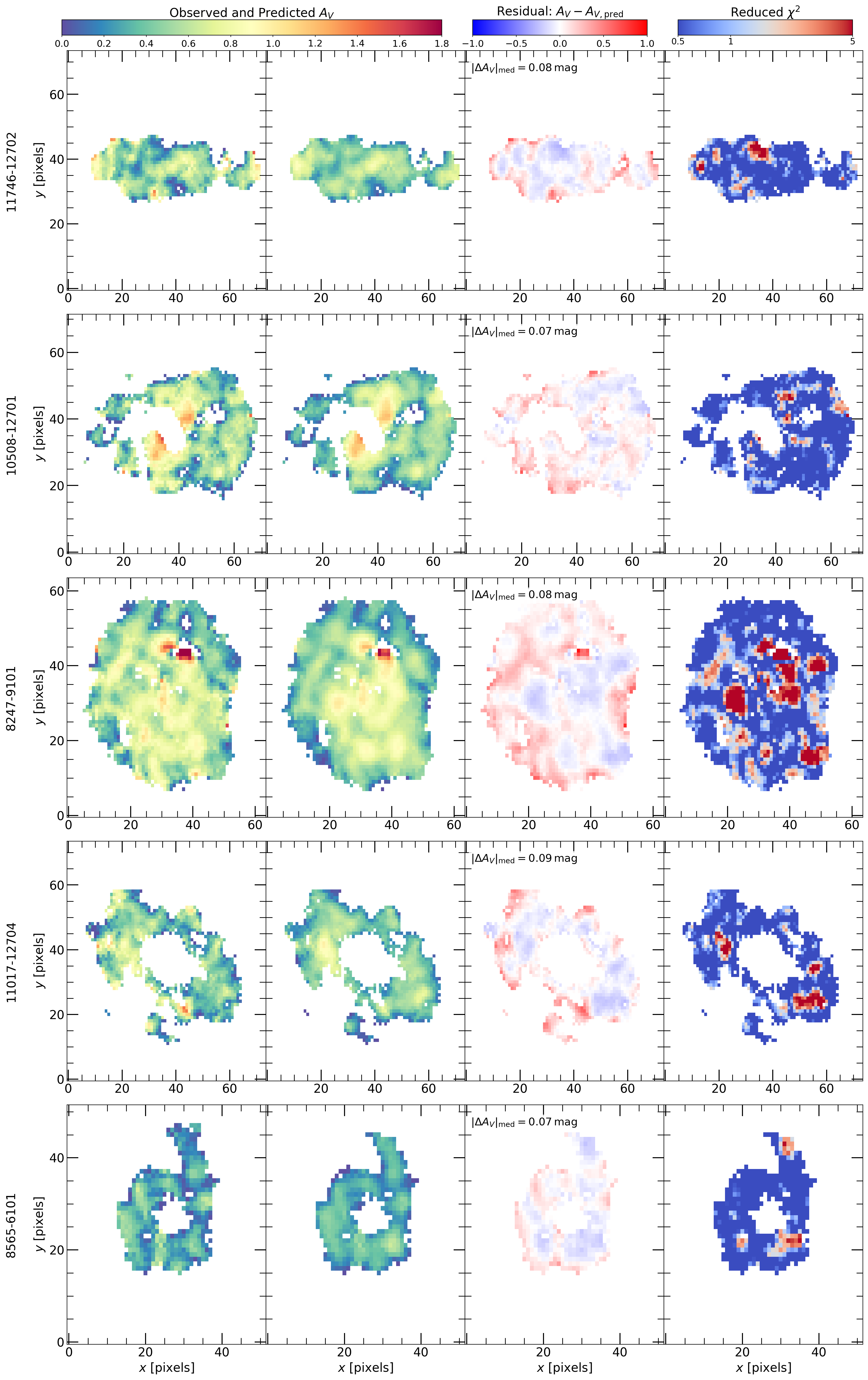}
    \caption{Comparison of observed and predicted $A_V$ maps for five representative MaNGA galaxies. Each row corresponds to a galaxy, and the columns show: (1) the observed $A_V$ derived from the BD, (2) the $A_V$ predicted by our model using $\Sigma_{\mathrm{SFR}}$, (3) the residual map annotated with the median $|\Delta A_V|$ for each galaxy, and (4) the $\chi^2$ map computed from residuals and observational uncertainties. The sample includes an edge-on galaxy, two bulge-dominated galaxies, and two disk-dominated galaxies.} 
    \label{fig:predicted_av_maps}
\end{figure*}

\subsection{Reconstructing galaxy dust attenuation maps}\label{observed_vs_modelled_R_Re}

In this section, we assess the performance of the empirical model in predicting $A_V$ on a spaxel-by-spaxel basis. This comparison evaluates how well the model reproduces the observed $A_V$ maps for individual galaxies, demonstrating its potential to generate spatially resolved $A_V$ estimates in the absence of emission line measurements.

\subsubsection{Comparison with Observed $A_V$ Maps}\label{individual_maps}

In Fig.~\ref{fig:predicted_av_maps}, we present a comparison between the observed and predicted $A_V$ maps for five representative MaNGA galaxies spanning a range of morphologies and inclinations. The sample includes one edge-on (11746-12702, axis ratio $\text{b/a}=0.29$), two bulge-dominated galaxies (10508-12701 and 8565-6101, $n=4.1$ and $n=3.1$), and two disk-dominated galaxies (8247-9101 and 11017-12704, $n=1.5$ and $n=2.3$). Galaxies are classified as bulge-dominated ($n>2.5$) or disk-dominated ($n<2.5$), with $n$ denoting the S\'ersic index as defined in \citet{Shen03}. These galaxies were selected to illustrate the model's performance across different morphologies and orientations.

Each row shows the observed $A_V$ map (first column), the predicted map from our empirical model (second column), the residual map (third column), and the reduced $\chi^2$ distribution (fourth column). The reduced chi-square is computed as $\chi^2_{\nu, i} = \left(A_{V,i}^{\mathrm{obs}} - A_{V,i}^{\mathrm{pred}}\right)^2 / \sigma_{A_{V,i}}^2$, where $A_{V,i}^{\mathrm{obs}}$ is the observed dust attenuation, $A_{V,i}^{\mathrm{pred}}$ is the model prediction, and $\sigma_{A_{V,i}}$ is the observational uncertainty. Since each spaxel is treated independently, the number of degrees of freedom is $\nu = 1$, making this reduced $\chi^2$ equivalent to a standard per-spaxel $\chi^2$.

The predicted $A_V$ maps reproduce the large-scale spatial structure of the observed maps, capturing central dust peaks and extended $A_V$ profiles across the disks. The residual maps indicate that differences between observed and predicted $A_V$ values are generally small and centred near zero, with mean absolute residuals of $0.09$-$0.12$ mag. For all five galaxies, these residuals are smaller than the typical observational uncertainties (0.14-0.20 mag), indicating that the scatter is consistent with measurement errors and demonstrating predictive accuracy close to the observational limit. Across the five galaxies, reduced chi-square values range from 0.52 to 2.14, with four galaxies exhibiting $\chi^2_\nu < 1$. Notably, the edge-on galaxy (11746-12702) achieves $\chi^2_\nu = 0.88$, indicating good agreement despite its challenging orientation, where $A_V$ is strongly affected by line-of-sight geometry and projection effects \citep{Tuffs_2004}. Residual variations between galaxies remain small and show no clear dependence on morphology or inclination, suggesting that the model performs consistently across diverse galaxy types and orientations.

When analysing residuals as a function of $R/R_e$, we find no significant bias in the central regions ($R/R_e < 0.5$). In the outskirts ($R/R_e > 1.0$), however, residuals are consistently higher in four of the five galaxies, with median values ranging from 0.05–0.09 mag, suggesting a mild underestimation of $A_V$ in the outer disks. The only exception is galaxy 8565-6101, which shows negligible residuals in both its central and outer regions. Overall, these results demonstrate that our model performs reliably across diverse galaxy types and orientations, with only minor deviations at large galactocentric radii.

\begin{figure*}[ht]
    \centering
    \includegraphics[width=\linewidth]{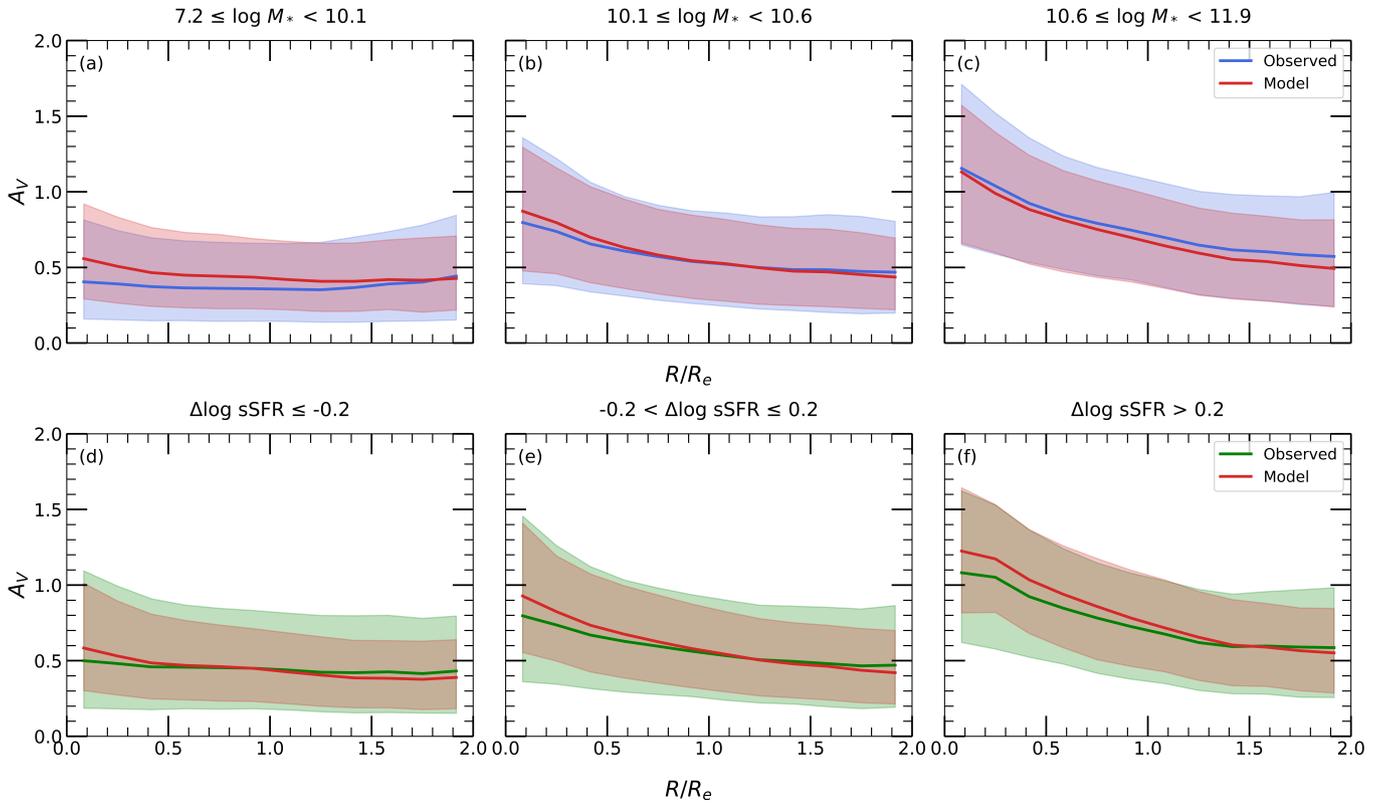}
    \caption{Radial profiles of observed and predicted $A_V$ across three global stellar mass bins (top row) and bins of global $\Delta \log\,\mathrm{sSFR}$ relative to the star-forming main sequence (bottom row). Solid lines show the median $A_V$ at each $R/R_e$, while shaded regions indicate the $1\sigma$ scatter. Predicted values from the model are shown in red in all panels. Observed $A_V$ is shown in blue (top row) and green (bottom row).} 
    \label{fig:radial_maps}
\end{figure*}

\subsection{Dust Attenuation Radial Trends Comparisons}\label{Radial_trends_comp}

\subsubsection{Dust Attenuation Radial Trends Comparisons on Global $M_*$}\label{Radial_trends_global_sm}

To evaluate the model’s performance, we test whether it reproduces systematic radial trends across the MaNGA sample. Radial gradients in $A_V$ are a well-established feature of star-forming galaxies, with attenuation typically peaking in the central regions and declining towards the outskirts. Previous studies have shown that the steepness and amplitude of these gradients depend on global $M_\ast$, with more massive galaxies exhibiting both higher central $A_V$ values and steeper radial declines \citep{greener20, Nelson_2016, Gonzalez_Delgado_2015}. This behaviour reflects the generally lower dust content in galaxy outskirts \citep{Casasola_2017, Draine2014}, consistent with the well-established radial decline in $\Sigma_\ast$ \citep{Elmegreen_2017, Grudic_2018} and the suppression of star formation in outer disks \citep{medling18}. To assess whether our model captures these trends, we compare the observed and predicted $A_V$ radial profiles across three $M_\ast$ bins: low ($7.2 < \log_{10}(M_\ast/M_\odot) < 9.7$), intermediate ($9.7 < \log_{10}(M_\ast/M_\odot) < 10.1$), and high ($10.1 < \log_{10}(M_\ast/M_\odot) < 11.14$), as shown in Fig.~\ref{fig:radial_maps}. The shaded regions denote the 16th–84th percentile range for each distribution.

We first examine the radial trends in the observed $A_V$ profiles across three $M_\ast$ bins. In the low-mass bin, the median $A_V$ is nearly flat, increasing slightly from $0.40$~mag in the center to $0.44$~mag in the outskirts. This corresponds to a mild increase of $+0.04$~mag and an average radial gradient of $0.02$~mag/$R_e$ (computed as $\Delta A_V / \Delta(R/R_e)$). The intermediate-mass bin shows a clear decline, with $A_V$ decreasing from $0.80$~mag in the center to $0.47$~mag in the outskirts, corresponding to a total drop of $0.33$~mag and a gradient of $-0.16$~mag/$R_e$. The steepest $A_V$ gradient is observed in the high-mass bin, where $A_V$ declines from $1.15$~mag at the center to $0.57$~mag in the outskirts, a total drop of $0.58$~mag and a gradient of $-0.29$~mag/$R_e$. These mass-dependent radial profiles indicate that dust is more centrally concentrated in massive galaxies. The steepening of $A_V$ gradients is linked to elevated central attenuation, sustained by the deeper gravitational potentials of massive systems that retain dust more effectively \citep{Goddard_2016, greener20}.

In the low-mass bin, the model reproduces the overall flatness of the observed profile but slightly overestimates the decline, with $A_V$ decreasing from $0.56$~mag in the center to $0.43$~mag in the outskirts ($\Delta A_V = -0.13$, gradient $=-0.07$~mag/$R_e$). In the intermediate-mass bin, the model slightly overestimates the steepness of the observed decline, predicting $A_V$ values of $0.87$~mag in the center and $0.44$~mag in the outskirts ($\Delta A_V = -0.44$, gradient $=-0.22$~mag/$R_e$), compared to the shallower observed gradient. In the high-mass bin, the model closely reproduces the steep observed decline, with $A_V$ decreasing from $1.13$~mag to $0.49$~mag ($\Delta A_V = -0.64$, gradient $=-0.32$~mag/$R_e$). Overall, the model reproduces both the mass dependence and radial steepening of the $A_V$ profiles, though it tends to predict slightly steeper gradients in the low and intermediate-mass bins, while accurately matching the strong decline in the most massive galaxies.

\subsubsection{Dust Attenuation Radial Trends Comparisons on Relative Global sSFR}\label{Radial_trends_global_sSFR}

We further examine $A_V$ radial profiles as a function of $\Delta \log \mathrm{sSFR}$ in Fig.~\ref{fig:radial_maps}, defined as the offset between a galaxy’s global $\log \mathrm{sSFR}$ and the $\,\mathrm{sSFR}$--$M_\ast$ relation from \citet{Grootes_2013}. This offset indicates whether a galaxy lies above or below the typical $\mathrm{sSFR}$–$M_\ast$ relation at fixed stellar mass, thereby serving as a measure of its relative star formation activity. To test this, we divide the sample into three $\Delta \log \mathrm{sSFR}$ bins: low ($\Delta \log \mathrm{sSFR} \leq -0.2$), intermediate ($-0.2 < \Delta \log \mathrm{sSFR} \leq 0.2$), and high ($\Delta \log \mathrm{sSFR} > 0.2$). Consistent with \citet{greener20}, we find that galaxies with higher relative sSFRs exhibit systematically higher overall $A_V$ values and show clear radial declines in $A_V$, which are not present in the lowest $\Delta \log \mathrm{sSFR}$ galaxy bin.

In the lowest bin, the observed $A_V$ profile remains almost flat, declining slightly from $0.50$~mag at the center to $0.43$~mag in the outskirts, a decrease of $-0.07$~mag with a gradient of $-0.03$~mag/$R_e$. Galaxies in the intermediate bin show a clearer decline, with $A_V$ decreasing from $0.80$ to $0.47$~mag, a decrease of $-0.33$~mag and a gradient of $-0.16$~mag/$R_e$. The steepest gradients occur in the highest bin, where $A_V$ falls from $1.08$ to $0.59$~mag across the radial range, corresponding to a decrease of $-0.50$~mag and a gradient of $-0.25$~mag/$R_e$. This is consistent with the findings of \citet{greener20}, who found that galaxies with higher relative star formation activity not only exhibit higher central attenuation, but also more pronounced radial declines in $A_V$. Such behaviour is expected from the well-established $\Sigma_\ast$–sSFR anticorrelation \citep{Castignani_20, Bezanson_2019}, where central regions with high $\Sigma_\ast$ sustain elevated star formation and dust production until molecular gas is depleted \citep{devis17, Elmegreen_1989}, while the outskirts maintain lower dust and star formation activity.

Similarly, we find that our model captures the radial attenuation trends across the $\Delta \log \mathrm{sSFR}$ bins. In the lowest bin, the model slightly overestimates the weak decline observed, predicting $A_V$ decreasing from $0.58$~mag to $0.39$~mag ($\Delta A_V = -0.19$, gradient $=-0.10$~mag/$R_e$). In the intermediate bin, it reproduces the higher central $A_V$ but predicts a steeper decline to $0.42$~mag ($\Delta A_V = -0.51$, gradient $=-0.25$~mag/$R_e$) compared with the observed. In the highest bin, the model again reflects the overall steepening but with a sharper drop, from $1.22$~mag to $0.55$~mag ($\Delta A_V = -0.67$, gradient $=-0.34$~mag/$R_e$). Overall, the model reproduces the increase in both central $A_V$ and gradient steepness with increasing relative sSFR, and although it tends to predict slightly stronger declines, its profiles remain consistent with the observations within the 1$\sigma$ scatter.

\section{Conclusions}\label{conclusion}

In this study, we developed and validated an empirical model to predict spatially resolved $A_V$ maps across star-forming galaxies using IFS data from the MaNGA survey at $0.0002 < z < 0.1444$.
Our analysis is based on 5,155 galaxies and $\simeq$1.9 million star-forming spaxels. We compared various combinations of physical parameters, including $\Sigma_\ast$, $\Sigma_{\text{SFR}}$, and $R/R_e$, to identify the optimal empirical model for predicting $A_V$. Our final model, based on an OLS regression using $\Sigma_{\text{SFR}}$, provides accurate predictions of $A_V$ across a broad range of local physical conditions. Our main findings are summarised as follows:

\begin{itemize}

    \item We find that $A_V$ can be reliably predicted from $\Sigma_{\text{SFR}}$, achieving $R^2$ = 0.69 and an RMSE of 0.22 mag, and recovering $A_V$ to within a factor of $\sim$1.3. The residuals are centred around zero, with a standard deviation of $\simeq0.2$ mag.

    \item $\Sigma_{\mathrm{SFR}}$ is the strongest predictor of $A_V$, with a Spearman rank correlation coefficient of $r = 0.80$, followed by $\Sigma_\ast$ ($r = 0.51$), while the normalised galactocentric radius $R/R_e$ shows only a weak negative correlation ($r = -0.11$). The limited role of $R/R_e$ reflects its dependence on the radial decline of $\Sigma_\ast$ and $\Sigma_{\mathrm{SFR}}$, reaffirming that local star formation activity is the dominant driver of attenuation, with $\Sigma_\ast$ providing secondary influence.

    \item Our empirical relation can be applied iteratively to recover dust–corrected $\Sigma_{\mathrm{SFR}}$ from attenuated fluxes. Applying our stopping criterion, the procedure converges by the fourth iteration, producing a minimal residual offset ($-0.01$ mag), a typical 1$\sigma$ scatter of $0.39$~dex, and a low RMSE of $0.42$~dex. This demonstrates that reliable $\Sigma_{\mathrm{SFR}}$ corrections can be achieved without BD measurements, extending the model's applicability to datasets lacking reliable H$\beta$ fluxes.

    \item After validating the model across the full MaNGA sample, we illustrate its performance using five representative galaxies spanning a range of morphologies and orientations. The model reproduces central dust peaks and extended radial profiles, with residuals near zero and typically within observational uncertainties. The performance is consistent across morphologies, including an edge-on system, with only mild underestimation of $A_V$ in outer regions ($R/R_e > 1.0$).

    \item Our model reproduces the $M_*$ and $\Delta \log \mathrm{sSFR}$ bin-dependent radial $A_V$ gradients, capturing the observed steepening of attenuation profiles in more massive galaxies and in those with higher relative star formation activity. While it slightly overpredicts central $A_V$ and underpredicts outer $A_V$ in these systems, the deviations remain within the 1$\sigma$ observational scatter.

\end{itemize}

We conclude that dust attenuation on kpc scales in star-forming regions of galaxies can be accurately predicted using local $\Sigma_{\text{SFR}}$. This model provides a computationally efficient complementary approach to radiative transfer models and offers a valuable tool for studying dust attenuation in spatially resolved galaxy surveys. In future work, this framework can be expanded by incorporating additional parameters, such as stellar age and gas-phase metallicity, to improve the accuracy of attenuation estimates across a broader range of galactic environments.

\section*{Data Availability}

We encourage authors to include a Data Availability Statement in their manuscript. This statement should include information on where resources such as data, materials, protocols and software code can be accessed. If data sharing is not applicable, authors should state that ‘Data sharing is not applicable to this article as no new data were created or analysed in this study.

%\begin{thebibliography}{}
%  \bibitem[\protect\citename{Akmajian and Lehrer, }1976]{akm76}
%   Akmajian \& Lehrer A. 1976, NP-like quantifiers and the
%   problem of determining the head of an NP. {\it Linguistic
%   Analysis\/} 11, 295--313.
%  \bibitem[\protect\citename{Huddleston, }1984]{hud84}
%   Huddleston, Rodney. 1984, {\it Introduction to the Grammar of
%   English}. Cambridge: Cambridge University Press.
%  \bibitem[\protect\citename{McCord, }1990]{mcc90}
%   McCord, Michael C. 1990, Slot grammar: a system for simpler
%   construction of practical natural language grammars. In R.
%   Studer (ed.), {\it Natural Language and Logic: International
%   Scientific Symposium}, pp.~118--45. Lecture Notes in Computer
%   Science. Berlin: Springer-Verlag.
%  \bibitem[\protect\citename{Salton {\it et al.}, }1990]{sal90}
%   Salton, Gerald, Zhao, Zhongnan \& Buckley, Chris. 1990,
%   A simple syntactic approach for the generation of indexing
%   phrases. Technical Report 90--1137, Department of Computer
%   Science, Cornell University.
%\end{thebibliography}

%\bibliography{magpi_dust}
\bibliographystyle{mnras}
\bibliography{magpi_dust}

\section{Acknowledgements}

Funding for the Sloan Digital Sky Survey IV has been provided by the Alfred P. Sloan Foundation, the U.S. Department of Energy Office of Science, and the Participating Institutions. SDSS-IV acknowledges support and resources from the Center for High Performance Computing at the University of Utah. The SDSS website is \href{https://www.sdss.org/}{www.sdss.org}.

SDSS-IV is managed by the Astrophysical Research Consortium of the Participating Institutions of the SDSS Collaboration including the Brazilian Participation Group, the Carnegie Institution for Science, Carnegie Mellon University, Center for Astrophysics | Harvard \& Smithsonian, the Chilean Participation Group, the French Participation Group, Instituto de Astrofísica de Canarias, The Johns Hopkins University, Kavli Institute for the Physics and Mathematics of the Universe (IPMU) / University of Tokyo, the Korean Participation Group, Lawrence Berkeley National Laboratory, Leibniz Institut für Astrophysik Potsdam (AIP), Max-Planck-Institut für Astronomie (MPIA Heidelberg), Max-Planck-Institut für Astrophysik (MPA Garching), Max-Planck-Institut für Extraterrestrische Physik(MPE), National Astronomical Observatories of China, New Mexico State University, New York University, University of Notre Dame, Observatário Nacional / MCTI, The Ohio State University, Pennsylvania State University, Shanghai Astronomical Observatory, United Kingdom Participation Group, Universidad Nacional Autónoma de México, University of Arizona, University of Colorado Boulder, University of Oxford, University of Portsmouth, University of Utah, University of Virginia, University of Washington, University of Wisconsin, Vanderbilt University, and Yale University.

%%%%%%%%%%%%%%%%%%%%%%%%%%%%%%%%%%%%%%%%%%%%%%%%%%
\section{Data Availability}

The data underlying this article were accessed from MaNGA Data Release 17. The DAP and \texttt{Pipe3D} pipeline are available at \href{https://www.sdss.org/dr17/manga/}{https://www.sdss.org/dr17/manga/}. Additional data generated by the analyses in this work are available upon request to the corresponding author.

%\appendix

\appendix

\input{appendix1.tex}

\end{document}

%% file: appendix1.tex
\section{}\label{appendix}

\begin{figure*}
    \centering
    \includegraphics[width=\linewidth]{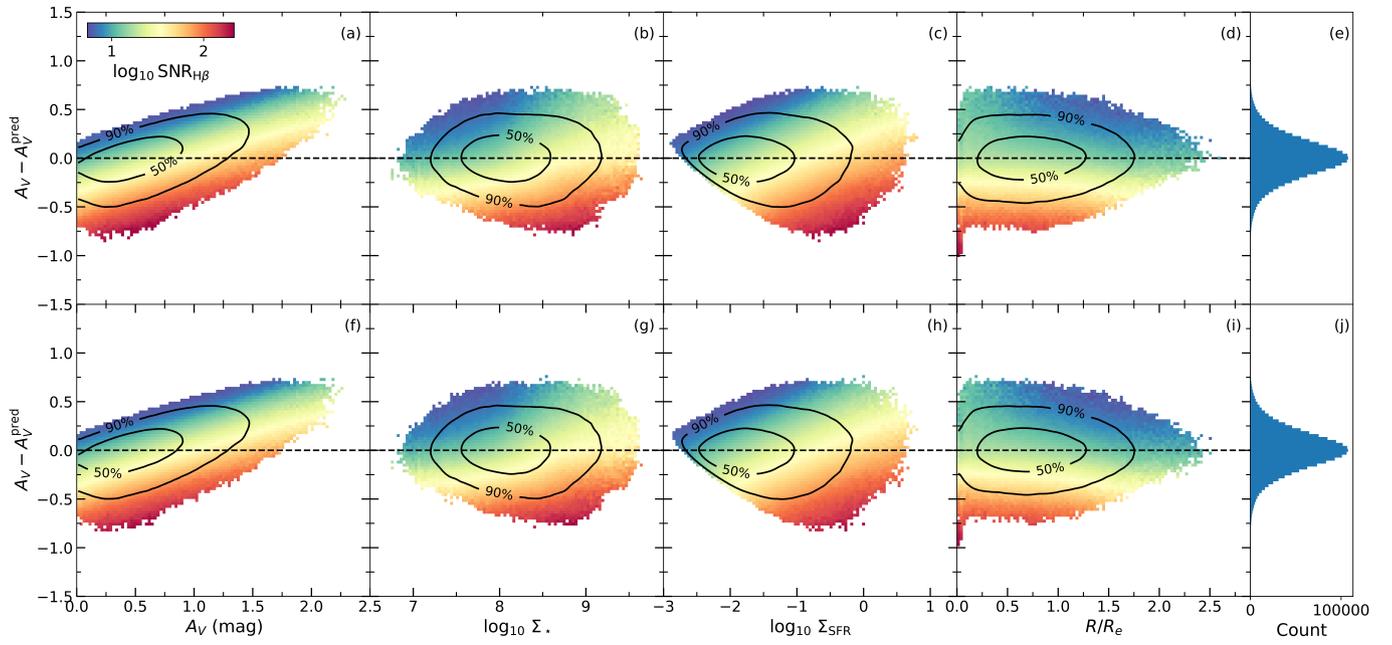}

    \caption{
        \textbf{Residual comparison between the 2D 
        ($\log\Sigma_{\mathrm{SFR}} + \log\Sigma_\ast$) 
        and 3D 
        ($\log\Sigma_{\mathrm{SFR}} + \log\Sigma_\ast + R/R_e$) 
        empirical $A_V$ models, shown in the top and bottom rows respectively.
        In each row, residuals are shown as a function of 
        observed $A_V$, $\log\Sigma_\ast$, $\log\Sigma_{\mathrm{SFR}}$, 
        and $R/R_e$, coloured by the median $\log_{10}(\mathrm{SNR}_{\mathrm{H}\beta})$ 
        in each bin. 
        Black contours mark the 50\% and 90\% spaxel density regions, 
        and the rightmost panels show the corresponding residual distributions as histograms.}
    }
    \label{fig:sfr_pred_appendix}
\end{figure*}